\documentclass[a4paper]{article}
\usepackage{ASVspoof}
\usepackage{epsfig,amssymb,amsmath}
\usepackage{hyperref}
\usepackage{booktabs,multirow}
\ninept
\usepackage{verbatim}
\usepackage{graphicx}
\usepackage{cite}

\title{USTC-KXDIGIT System Description for ASVspoof5 Challenge}

\makeatletter
\def\name#1{\gdef\@name{#1\\}}
\makeatother
\name{{\em Yihao Chen$^{1,2}$\textsuperscript{*}, Haochen Wu$^{1,2}$\textsuperscript{*}, Nan Jiang$^{1,2}$, Xiang Xia$^{1}$, Qing Gu $^{1,2}$ } \\
{\em Yunqi Hao$^{1,3}$, Pengfei Cai$^{1,2}$,  Yu Guan$^{2}$, Jialong Wang$^{1,3}$,  Weilin Xie$^{1,2}$} \\
      {\em Lei Fang$^{1}$,  Sian Fang$^{1}$, Yan Song$^{2}$, Wu Guo$^{2}$, Lin Liu$^{1}$, Minqiang Xu$^{1}$\textsuperscript{\dag}}}

\address{$^1$Hefei iFly Digital Technology Co.Ltd, Hefei, China  \\
$^2$University of Science and Technology of China, Hefei, China  \\
$^3$Xinjiang University, Urumqi, China  \\
{\small \tt miracle1025@mail.ustc.edu.cn} }
\begin{document}
\maketitle
\begingroup\renewcommand\thefootnote{*}
\footnotetext{These authors share equal contribution to this work.}
\begingroup\renewcommand\thefootnote{\dag}
\footnotetext{Corresponding author.}
\begin{abstract}
This paper describes the USTC-KXDIGIT system submitted to the ASVspoof5 Challenge for Track 1 (speech deepfake detection) and Track 2 (spoofing-robust automatic speaker verification, SASV). Track 1 showcases a diverse range of technical qualities from potential processing algorithms and includes both open and closed conditions. For these conditions, our system consists of a cascade of a front-end feature extractor and a back-end classifier. We focus on extensive embedding engineering and enhancing the generalization of the back-end classifier model. Specifically, the embedding engineering is based on hand-crafted features and speech representations from a self-supervised model, used for closed and open conditions, respectively. To detect spoof attacks under various adversarial conditions, we trained multiple systems on an augmented training set. Additionally, we used voice conversion technology to synthesize fake audio from genuine audio in the training set to enrich the synthesis algorithms. To leverage the complementary information learned by different model architectures, we employed activation ensemble and fused scores from different systems to obtain the final decision score for spoof detection. During the evaluation phase, the proposed methods achieved 0.3948 minDCF and 14.33\% EER in the close condition, and 0.0750 minDCF and 2.59\% EER in the open condition, demonstrating the robustness of our submitted systems under adversarial conditions. In Track 2, we continued using the CM system from Track 1 and fused it with a CNN-based ASV system. This approach achieved 0.2814 min-aDCF in the closed condition and 0.0756 min-aDCF in the open condition, showcasing superior performance in the SASV system.
\end{abstract}

\section{Introduction}
Automatic Speaker Verification (ASV) systems has been widely used in many human-machine interfaces, where a person’s identity is automatically verified in biometric authentication. Meanwhile, with recent advances in deep learning based text-to-speech (TTS) and voice conversion (VC), speech synthesis \cite{1} can readily generate extremely near-realistic and human-like speech, which however is often indistinguishable to human ears and poses a serious threat to state-of-the-art ASV systems. Therefore, to deal with these issues, it is urgent to develop effective and robust spoof speech detection systems.
%TTS and VC techniques like WaveNet [1], Deep Voice [2] and Tacotron [3] greatly enhanced the quality of the voice-spoofed utterances

In practice, speech deepfake detection has drawn great attention in the AI community and the technology industry. To benchmark the progress of research in speech deepfake detection, the ASVspoof Challenge \cite{2,3,4,5} challenge releases a series of spoofing datasets to  imitate diversified and challenging attack situations in realistic applications, including TTS, VC and speech replay attacks. The ASVspoof 2021 challenge \cite{5} has included more data to simulate quite practical and realistic scenarios of different spoofing attacks. There are three sub-challenges: physical access (PA), logical access (LA) and deepfake detection (DF). LA and DF both aim to detect the spoofing speech generated by various advanced text-to-speech (TTS) and voice conversion (VC) techniques. The ASVspoof 2021 LA aimed to evaluate the robustness of the spoofing detection model against different channel effects by adding various codec and channel transmission effects to the evaluation data. As a extended from the LA track, the ASVspoof 2021 DF task aims to evaluate the spoofing detection system against different unknown conditions. In addition to the series of ASVspoof challenges, the Audio Deepfake Detection Challenge (ADD) \cite{6} extends the attack situations of fake audio detection to further address diversified and challenging attack situations in realistic applications. To address the issue of voice spoofing, the first Spoofing-Aware Speaker Verification (SASV) Challenge \cite{25} aims to integrate research efforts in speaker verification and anti-spoofing.
%Therefore, the detection systems for both LA and DF tracks need to be robust to unseen attacks and audio compression techniques.

Recently, self-supervised pre-trained model has achieved significant advances in the fields of speech deepfake detection\cite{7,8,9}, which far exceeds the performance of conventional methods. It was shown that building a general pre-trained model based on the exploitation of a large mount of unlabeled data can be quite essential to boost the performance, reduce data labeling efforts and lower entry barriers for speech deepfake detection. In this work, we also apply the open-sourced wav2vec2-large pre-trained model as the feature extractor to help build a robust detection system. However, deepfake speech exhibits local/global artifacts \cite{10}, which vary across different TTS and VC algorithms, and speech could be of varied technical quality stemming from the application of different processing algorithms. These both imply the impossibility of being fully covered by the training set of an SSD system. Generally, the most intuitive way to address the abovementioned issues is data augmentation \cite{9,11,12}. Another solution is to provide multi-scale information for model training \cite{13,14}, which can avoid information loss in the training process by increasing model capacity. In addition, a potential direction for improving generalization ability is one-class learning. OC-softmax and its variants \cite{15,16} were proposed for speech deepfake detection, which builds the feature space by employing two margins for genuine and spoofed speech.

%This paper presents our submitted system to the spoof5spoof5 Challenge. Based on these aforementioned observations, we experimented with several spoof detection systems. First, we trained the systems with training data augmented with noises and diversified processing algorithms, aiming to enable the systems to detect the spoof artifacts under severe conditions. Moreover, to better obtain deepfake artifacts, we explore the use of different neural network architectures, and the fusion of hand-craft features and pre-trained model embedding. Furthermore, to exploit the complementarity of different architectures in terms of spoof detection, we have performed activation ensemble technique and weighted score-level fusion to obtain the final CM scores. For the track2, we first trained a CNN-architecture ASV network, utilizing a two-stage training method to complete fine-tuning, and used QMF ackend techniques to optimize the ASV scores.Then we continued using the CM system from Track 1 and use cascaded method to fuse it with the ASV system.
This paper presents our system submitted to the ASVspoof5 \cite{43} Challenge. Building on the observations mentioned, we experimented with several spoof detection systems. First, we trained the systems using augmented training data with added noises and various processing algorithms to help detect spoof artifacts under challenging conditions. Additionally, to better capture deepfake artifacts, we explored different neural network architectures and the fusion of handcrafted features with pre-trained model embeddings. To leverage the complementary strengths of different architectures in spoof detection, we implemented activation ensemble techniques and weighted score-level fusion to derive the final CM scores. For Track 2, we initially trained a CNN-based ASV network, using a two-stage training method for fine-tuning, and applied QMF backend techniques to optimize the ASV scores. We then continued using the CM system from Track 1, employing a cascaded method to integrate it with the ASV system.

The remainder of this paper is organized as follows. Section 2 introduces the challenge task and datasets. Section 3 and 4 describe the augmentation method and system architectures in detail, respectively. Section 5 introduces the experimental results, followed by concludes in Section 6.

\section{Task Description and Datasets}
\subsection{Task Description}
ASVspoof5 \cite{43} is centred around two subtasks: stand-alone speech deepfake detectors (Track 1), and spoofing-robust automatic speaker verification solutions (Track 2). There are two conditions, open and closed, for both Track 1 and Track 2. For closed condition, participants commit to using data within the ASVspoof5 training partition only, and however participants may use external data and pre-trained models in the open condition. 
%Therefore, or better assessment of countermeasures for various spoofing attacks, the following problems should be considered:
%\vspace{-5mm}
%\begin{itemize}
%\itemsep -1.3mm
%\item The speech samples in the evaluation set will be of varied technical quality stemming from the potential application of speech coding, audio compression, bandlimiting or other processing algorithms.
%\item We can use external data and pre-trained models to develop robust spoof speech detection systems.
%\end{itemize}

\subsection{Datasets}
For the Track 1 Closed task, we only used the training data from ASVspoof5. For Track 1 Open task , the datasets used for training included:
\begin{itemize}
\itemsep -1.3mm

\item ASVspoof5 train data

\item ASVspoof2021 LA,ASVspoof2021 DF data

\item Our own business data
\end{itemize}
For the Track 2 tasks, our CM system remained consistent with Track 1, while the ASV system used only the VoxCeleb2 data.
%The contents of the ASVspoof5 datasets are summarized in Table 1. The challenge itself ran in two phases: progress phase and evaluation phase. The progress dataset is a subset of the evaluation data, which help participant to know the performance of systems. During the evaluation phase, participants will be allowed only a single submission. Evaluation was performed in terms of minimum Tandem Detection Cost Function (min-tDCF) \cite{17} as primary metric.

\begin{comment}
\begin{table}[th]
  \centering
  \caption{\it Final submission system for Track1 open condition.}
  \label{tab:1}
  \renewcommand\arraystretch{1.2}
  \setlength{\tabcolsep}{3.5pt}{
  \begin{tabular}{cccccc}
    \toprule
    %\toprule[1.3pt]
    \multirow{2}{*}{Data subset} 
    &\multicolumn{2}{c}{\# Speakers}
    &\multicolumn{3}{c}{\# Utterances}\\   
    \cline{2-6}
    \specialrule{0em}{0.8pt}{1pt}  
       ~ &Male  &Female &Bonafide  &Spoof &Total\\
    \midrule
       Training    &- &- &- &- &182,358\\   
       Development &- &- &- &- &140,950\\
       Progress    &- &- &- &- &40,765\\
       Evaluation  &- &- &- &- &680,774\\  
    \bottomrule
    %\bottomrule[1.3pt]
  \end{tabular}}
\end{table}
\end{comment}

\section{Augmentation techniques}
\subsection{CM system}
\subsubsection{Online augmentation}

The number of utterances in the evaluation data set was much larger than in training or developing sets. Therefore, to reduce over-fitting and bias caused by diversified noise, we apply different augmentation methods for different conditions to increase system robustness.

For the closed condition,  we apply a variety of codec
algorithms \cite{5}, including MP3, OGG, AAC OPUS, a-law
and µ-law. Besides, to mock the telephony transmission
loss, audio samples are first downsampled to 8kHz and
then upsampled back to 16kHz.

For the open condition, first, we utilized recorded noises from MUSAN \cite{18} and the RawBoost method \cite{11} to add noises. Based on a variety of convolutional and additive noises, RawBoost models nuisance variability stemming such as encoding, transmission, microphones and amplifiers, as well as linear and nonlinear distortions. Thus we can add different nuisance noises dependent on the corresponding raw waveform inputs. Second, we randomly selected audio from the public room impulse response (RIR) set \cite{19} and convoluted it with the target audio to generate new audio for reverberation simulation of different room sizes. Besides, we apply a variety of codec algorithms \cite{5}, including MP3, OGG, AAC, OPUS, a-law, µ-law and so on. The training data of all models in the experiment were augmented using the above-mentioned methods in the on-the-fly way.

%\subsubsection{Waveform augmentation}

%We used online augmentation to expand the training data to increase system robustness.

\subsubsection{Synthesis algorithms augmentation}
Most DA methods focus on improving the generalization of the system in real scenes, such as the methods
in Section 3.1, which, however, cannot cover the speech
synthesis algorithms in the training set. To tackle this
issue, we use several vocoders to synthesize fake audio. In this work, we select three neural vocoders for the closed condition and one neural vocoder for the open condition. We utilized the bona fide speech from the ASVspoof 5 training dataset to train the FastDiff\cite{26}, FreeV\cite{27}, and HiFi-GAN\cite{28} models for the closed condition. For the open condition, we used the VCTK corpus to train the VITS\cite{29} model.

\textbf{HiFi-GAN} : As one of the state-of-the-art neural vocoders, HiFi-GAN generates audio based on
generative adversarial networks (GANs) using
the true mel-spectrum. It includes one generator
and two discriminators: multi-scale and multi-period discriminators, which can achieve efficient
and high-fidelity speech synthesis. The generator
and discriminators are trained adversarially by
incorporating two additional losses to improve
the training stability and model performance.

\textbf{FastDiff} : FastDiff is a speedy conditional diffusion model designed for high-quality speech synthesis. To enhance audio quality, FastDiff employs a series of time-aware location-variable convolutions with various receptive field patterns. This approach efficiently models long-term temporal dependencies using adaptive conditions.

\textbf{FreeV} : A vocoder framework that simplifies the model's predictive complexity by utilizing the estimated amplitude spectrum. This vocoder is composed of PSP and ASP, with ConvNeXtV2 serving as the foundational building block. The PSP features an input convolutional layer, eight ConvNeXtV2 blocks, and two convolutional layers for a parallel phase estimation structure.

\textbf{VITS} : The VITS model includes a text encoder that converts text into linguistic features, a flow-based posterior encoder that captures the complex distribution of these features, and a HiFi-GAN-based decoder that synthesizes the final audio waveform. By leveraging these components, VITS is capable of producing high-quality speech with rich prosody and naturalness, making it a powerful tool for text-to-speech applications. We use the VCTK corpus to train the VITS model. During inference, the ASR (Automatic Speech Recognition) results from a LibriSpeech-based model and the speaker representation extracted from ASVspoof5 training data are used to decode the waveform.

\subsection{ASV system}
For both closed and open condition, we only use the VoxCeleb2\cite{30} dataset to perform system development. We here adopted a 3-fold speed augmentation at first to generate extra twice speakers. Each speech
segment in this dataset was perturbed by 0.9 and 1.1 factors
based on the SoX speed function. Then we obtained total 17982
speakers, which is triple amount of the original speakers.
We also applied the following techniques to augment each utterance:

• Reverberation: artificially reverberation using a convolution with simulated RIRs from the AIR dataset

• Noise: MUSAN noises were added at one-second intervals throughout the recording (0-15dB SNR)
%\subsubsection{Headings}
%Section headings are centered in boldface
\section{Systems Description}
\subsection{CM system}
%\subsubsection{Embedding extraction for closed condition}
\subsubsection{Single models for closed condition}
We used the baseline-based AASIST and RawNet2\cite{35} networks. Furthermore, we employ the S$^2$pecNet\cite{33} method for audio spoofing detection, which harnesses complementary information from multi-order spectrograms. This approach features a TSF module that merges the two spectral representations in a coarse-to-fine manner. To minimize information loss, the fused representation is reconstructed back into the original spectrograms. On the other hand, we used AMSoftmax \cite{31} and Circle Loss \cite{32} in place of the standard Softmax loss function for training. Additionally, we experimented with using different fixed-length of durations for inference and then fused the results.
\subsubsection{Fusion for closed condition}
To construct final system for closed condition, we used 4 single subsystems. Fusion was performed on the score level.
The optimal weights are estimated using COBYLA \cite{34} toolkit
minimizing the minDCF metric on development set.

\begin{table}[]
\caption{\label{table1} {\it Final submission system for Track1 closed condition.}}
\vspace{2mm}
\begin{tabular}{|l|l|l|ll}

\cline{1-3}
classifer               & Feature                    & Loss Function &  &  \\ \cline{1-3}
\multirow{2}{*}{AASIST} & \multirow{2}{*}{Raw audio} & AMSoftmax     &  &  \\ \cline{3-3}
                        &                            & Circle loss   &  &  \\ \cline{1-3}
S$^2$pecNet                & Raw audio \& LFCC          & AMSoftmax     &  &  \\ \cline{1-3}
RawNet2                  & Rawaudio                   & Softmax       &  &  \\ \cline{1-3}
\end{tabular}
\end{table}
\subsubsection{Embedding extraction for open condition}

Wav2vec2 \cite{20} is a self-supervised pre-trained model, which can extract speech representations or embeddings from raw waveform. It has shown an impressive performance on speech deepfake detection. Therefore, we use Wav2vec2-large as the feature extractor for the open condition. 

Fine-tuning Wav2vec2-large model mainly includes three stages. Firstly, the raw waveform is sent into a feature encoder composed of several convolutional layers (CNN). The feature encoder extracts vector representations of size 1024 every 20ms and the receptive field is 25ms. Secondly, these encoder embeddings are fed into the context encoder, which contains 24 transformer block layers and is used to explore the contextual information contained in the input speech. Finally, the context encoder embeddings are used for downstream task.

Even though the features extracted by Wav2vec2 can include rich discriminative spoofing information during fine-tuning process, they still have a tendency to overlook some principles of speech pronunciation or hearing. On the contrary, Linear frequency cepstral coefficients (LFCCs) obtains great detection performance in the hand-craft features, which reflects LFCCs are data-independent and robust even in the face of unknown attacks \cite{21}. Therefore, it would be interesting to see if leveraging the complementarity of these two sets of features can help SDD.

\subsubsection{Single models for open condition}
AASIST \cite{22} is an end-to-end system using Integrated Spectro-Temporal graph attention networks, which mainly includes three innovations: 1) a novel heterogeneous stacking graph attention layer, which models artefacts spanning heterogeneous temporal and spectral domains with a heterogeneous attention mechanism and a stack node, 2) a max graph operation  that involves a competitive selection of artefacts, and 3) a modified readout scheme. AASIST uses a sinc convolutional layer based front-end, and thus can extract representations directly from raw waveform inputs. To integrate AASIST with self-supervised model when fine-tuning, the sinc convolution layer of AASIST is replaced by the aforementioned wav2vec2 model as the same model architecture as in \cite{8}.

Meanwhile, there exists a diversity in the spoof utterances
generated by different text-to-speech and voice conversion algorithms, resulting in a poor generality of an SSD system
to unseen spoofing attacks. To address this problem, we adopt res2net-based system as the same model architecture as in \cite{9}, which integrate multi-scale feature aggregation (MFA) and dynamic convolution operations into the anti-spoofing framework. This additional multi-scale reception improves the system’s capacity and helps the system perform better when generalized to unseen spoofing attacks

Besides, to leveraging the complementarity of LFCCs and wav2vec2 embedding, we apply the aggregation methods as in \cite{23}.
The wav2vec2 process each speech segment with a 20ms frame-shift,
whereas LFCCs utilizes a 10ms frame-shift. Therefore, it is crucial to align the two features in prior to being applied to aggregation modules. Specifically, the extracted 60-dimensional LFCCs features are first processed using Conv1d to align both the time and feature dimensions with wav2vec2 features.
\begin{table}[ht]
\vspace{-1em} 
\caption{\label{table1} {\it Final submission system for Track1 open condition.}}
\vspace{2mm}
\centerline{
\begin{tabular}{|c|c|c|}
%\begin{tabular}{|c|c|}
\hline
Classifier & Feature & Training data \\
\hline
%\hline  \hline
\multirow{4}{*}{AASIST} & \multirow{4}{*}{Wav2vec2 embedding} & Ori \\
\cline{3-3}
~   & ~ & Add LA \\
\cline{3-3}
~   & ~ & Add DF \\
\cline{3-3}
~   & ~ & Add business data \\
\hline
\multirow{4}{*}{MFA-Res2net} & \multirow{4}{*}{Wav2vec2 embedding} & Ori \\
\cline{3-3}
~   & ~ & Add LA \\
\cline{3-3}
~   & ~ & Add DF \\
\cline{3-3}
~   & ~ & Add business data \\
\cline{3-3}
\hline
\multirow{4}{*}{MFA-Res2net} & Wav2vec2 embedding & Ori \\
\cline{3-3}
~   & \multirow{2}{*}{\&} &  Add LA \\
\cline{3-3}
~   & ~ & Add DF \\
\cline{3-3}
~   & LFCC & Add business data \\
\hline
\end{tabular}}
\end{table}
\subsubsection{Training Protocol}
In the closed condition, during the training stage, the speech is cropped or concatenated to a fixed length of 4 seconds. We configure the sinc layer with 70 filters, each having fixed cut-in and cut-off frequencies. The Adam optimizer, with a learning rate of $8 \times 10^{-4}$, is utilized. The training batch size is set to 32. In the test phase, we use different fixed durations (2 seconds, 4 seconds, and 6 seconds) and fuse the results at the score level.

For open condition, all the single models trained were based of wav2vec2 embedding extractor. Besides, a fully connected layer after the pre-trained model is used to reduce the representation dimension from 1024 to 256. During fine-tuning, the pre-trained wav2vec2 model is optimized jointly with these back-end network via the back-propagation. We use the standard Adam optimizer , which adopts a mini-batch size of 16 and a learning rate of $10^{-6}$ with a weight decay of $10^{-4}$ to avoid over-fitting. Considering the imbalance between the genuine and fake audios in the training set, we use the weighted cross entropy to minimize the training loss. The weights are associated with the number of the genuine and fake categories.
\subsubsection{Fusion for open condition}
To construct final system for open condition, we used 12 single subsystems which can be divided into three parts according to
the architecture they are based on (Table 3). These are 4 Light
Res2Net-like, 4 AASIST-like models and 4 Res2Net-LFCCs models. The details of final system is demonstrated in Table 1. Fusion was performed on the score level. Fusion weights were selected manually with respect to the performance of each single system on the progress subset of evaluation set.
\subsection{ASV system}
\subsubsection{Backbone}
As one of the most classical ConvNets, ResNet \cite{36} has proved
its power in speaker verification. In our systems, bottleneckblock-based ResNet (deeper structures:ResNet-242) are adopted. Base channels of all these ResNets
are 64. Similar to the classic ResNet, the network structure begins with a 3×3 convolution layer as the initial layer, followed by four convolution blocks. Each block contains stacked bottleneck layers. The strides for these blocks are (1, 2, 2, 2), and the number of channels in each block is (64, 128, 256, 512). The number of bottlenecks in each block of ResNet-242 is (3, 10, 64, 3).
\subsubsection{Pooling method}
The pooling layer aims to aggregate the variable sequence to an utterance level embedding. We used the multi-query multi-head attention pooling mechanism layer (MQMHA)\cite{37} in the ResNet242 model. The number of query was set to 4 while the number
of head was 16.
\subsubsection{Loss function}
Recently, margin based softmax methods have been widely used
in speaker recognition works. To make a much better performance, we strengthen the AM-Softmax and AAM-Softmax\cite{39}  loss functions by Sub-Center \cite{38} method.
\subsubsection{Backend}
Quality Measure Functions (QMF) \cite{40} was applied to calibrate
the scores, and it greatly enhanced the performance. For QMF,
we combined two qualities, speech duration and magnitude of embeddings. We selected dev trials from the asvspoof5 
as the training set of QMF. Then a Logistic Regression (LR) was
trained to serve as our QMF model.
For speech duration, we used duration of enroll and test as
quality. We used QMF to calibrate the system score.
\subsubsection{Training Protocol}
We conducted our experiments using PyTorch, training the ASV model in two stages. In the first stage, we used the SGD optimizer with a momentum of 0.9 and a weight decay of 1e-3. The models were trained on 8 GPUs with a mini-batch size of 1,024 and an initial learning rate of 0.08. For each batch, 200 frames per sample were used. The ReduceLROnPlateau scheduler was employed, validating every 2,000 iterations with a patience of 2. The minimum learning rate was set to 1.0e-6, and the decay factor was 0.1. Additionally, the margin was gradually increased from 0 to 0.2. We utilized the SubCenter loss function in this stage, with the number of SubCenters (K) set to 3.

In the Large-Margin Fine-Tuning stage (LM-FT)\cite{41}, settings are slightly different from the first stage. First, we removed the speed augmentation from the training set to avoid domain mismatch. Next, we changed the frame size from 200 to 600 and increased the margin exponentially from 0.2 to 0.8. The AM-Softmax loss was replaced with AAM-Softmax loss. Third, we selected the center with the largest norm as the "dominant center" and discarded the other sub-centers, using only the dominant center as the speaker center throughout the fine-tuning process. Finally, we used a smaller fine-tuning learning rate of 8e-5 with a batch size of 256. The learning rate scheduler remained mostly unchanged, except the decay factor was adjusted to 0.5.

\begin{figure}[htbp]
\vspace{-1em} 
\centering
\includegraphics[width=0.5\textwidth]{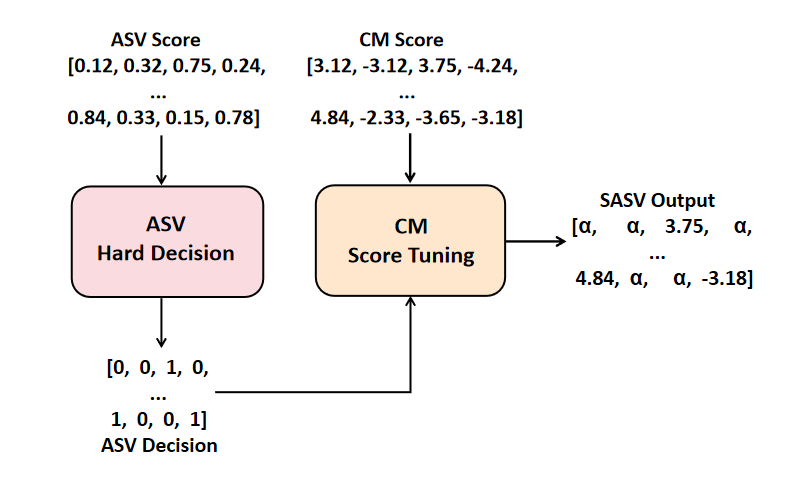}
\caption{ The illustration of the ASV followed by CM cascaded system. $\alpha$ denotes the minimum CM score in the development set.}
\label{fig:myfigure}
\vspace{-1em} 
\end{figure}

\subsubsection{Fusion for CM and ASV system}
We use a cascaded system to integrate the CM (countermeasure) and ASV (automatic speaker verification) systems. The cascaded system is composed of two tandem modules:

a) The first module generates a hard decision based on a threshold, which is set according to the a-DCF on the development set.

b) If the decision from the first module is positive, the second module directly outputs the raw scores.

c) If the decision from the first module is negative, the second module assigns a fixed minimum score based on the development set.

Thus, if the test audio is flagged as negative by the first module, the score from the second module is disregarded.

\section{Results and Discussion}
\subsection{ASVSpoof5 Track1}
    \begin{table}[h!]
    \vspace{-1em} 
    \setlength{\abovecaptionskip}{-3pt}
    \centering
    \footnotesize
    \caption{\raggedright {\it Results on Track1 closed condition.}}
    \vspace{2mm}
    \label{tab:Track1 closed_condition}
    \begin{tabular}{|c|c|c|c|c|c|c|}
        \hline
        & A17 & A18 & A19 & A20 & A21 & A22 \\
        \hline
        minDCF & 0.4209 & 0.3740 & 0.6255 & 0.3531 & 0.2456 & 0.3257 \\
        EER(\%) & 14.80 & 13.02 & 31.80 & 12.19 & 9.42 & 11.65 \\
        \hline
        & A23 & A24 & A25 & A26 & A27 & A28 \\
        \hline
        minDCF & 0.2726 & 0.2800 & 0.1827 & 0.3531 & 0.3773 & 0.8381 \\
        EER(\%) & 10.08 & 10.94 & 6.78 & 12.18 & 13.81 & 28.99 \\
        \hline
        & A29 & A30 & A31 & A32 & \multicolumn{2}{c|}{pooled}\\
        \hline
        minDCF & 0.2307 & 0.4443 & 0.4355 & 0.4066 & \multicolumn{2}{c|}{0.3948} \\
        EER(\%) & 8.63 & 16.02 & 16.09 & 14.51 & \multicolumn{2}{c|}{14.33} \\
        \hline
    \end{tabular}

\end{table}
The final results of the evaluation set system submitted for the Track 1 closed challenge are presented in Table 3. Notably, the A19 and A28 spoofing algorithms underperformed, yielding a final minDCF of 0.3948 and an EER of 14.33\%.

\begin{table}[h!]
    \vspace{-1em} 
    \centering
    \footnotesize
    \caption{\raggedright {\it Results on Track1 open condition.}}
    \vspace{2mm}
    \label{tab:Track1 Open condition}
    \begin{tabular}{|c|c|c|c|c|c|c|}
        \hline
        & A17 & A18 & A19 & A20 & A21 & A22 \\
        \hline
        minDCF & 0.0108 & 0.0502 & 0.0593 & 0.0733 & 0.0163 & 0.0248 \\
        EER(\%) & 0.39 & 1.80 & 2.06 & 2.55 & 0.57 & 0.86 \\
        \hline
        & A23 & A24 & A25 & A26 & A27 & A28 \\
        \hline
        minDCF & 0.0527 & 0.0080 & 0.0166 & 0.0288 & 0.0949 & 0.2218 \\
        EER(\%) & 1.86 & 0.28 & 0.59 & 1.15 & 3.47 & 9.62 \\
        \hline
        & A29 & A30 & A31 & A32 & \multicolumn{2}{c|}{pooled}\\
        \hline
        minDCF & 0.0128 & 0.0818 & 0.0784 & 0.0725 & \multicolumn{2}{c|}{0.0750} \\
        EER(\%) & 0.48 & 2.84 & 2.72 & 2.67 & \multicolumn{2}{c|}{2.59} \\
        \hline
    \end{tabular}

\end{table}
The final results of the evaluation set system fusion submitted for the Track 1 open challenge are shown in Table 4. With the enhancement of self supervised pre training models, the detection performance of various spoof algorithms has improved significantly , achieving a final min a-DCF of 0.0750.

\subsection{ASVSpoof5 Track2}

\begin{table}[h!]
   \vspace{-1.5em} 
    \centering
    \footnotesize
    \caption{\raggedright {\it Results on Track2 closed condition.}}
    \vspace{2mm}
    \scalebox{0.95}{
    %\label{tab:Track2 closed_condition}
    \begin{tabular}{|c|c|c|c|c|c|c|}
        \hline
        & A17 & A18 & A19 & A20 & A21 & A22 \\
        \hline
        min a-DCF & 0.3220 & 0.2667 & 0.5087 & 0.2461 & 0.1560 & 0.2202 \\ % 替换了这些数值
        
        \hline
        & A23 & A24 & A25 & A26 & A27 & A28 \\
        \hline
        min a-DCF & 0.1784 & 0.1747 & 0.0839 & 0.2060 & 0.2596 & 0.6304 \\ % 替换了这些数值        
        \hline
        & A29 & A30 & A31 & A32 & \multicolumn{2}{c|}{pooled}\\
        \hline
        min a-DCF & 0.1464 & 0.3277 & 0.3152 & 0.2697 & \multicolumn{2}{c|}{0.2814} \\ % 替换了这些数值        
        \hline
    \end{tabular}
}
\end{table}

The final results of the fused evaluation set system submitted for the Track 2 closed challenge are presented in Table 5. Influenced by the CM system from Track 1, the A19 and A28 algorithms also performed poorly in the Closed mode of Track 2, resulting in a final min a-DCF of 0.2814.

\begin{table}[h!]
    \vspace{-1em} 
    \centering
    \footnotesize
    \caption{\raggedright {\it Results on Track2 open condition.}}
    \vspace{2mm}
    \label{tab:Track2 Open condition}
    \scalebox{0.95}{
    \begin{tabular}{|c|c|c|c|c|c|c|}
        \hline
        & A17 & A18 & A19 & A20 & A21 & A22 \\
        \hline
        min a-DCF & 0.0393 & 0.0586 & 0.0666 & 0.0721 & 0.0409 & 0.0445 \\       
        \hline
        & A23 & A24 & A25 & A26 & A27 & A28 \\
        \hline
        min a-DCF & 0.0608 & 0.0371 & 0.0382 & 0.0437 & 0.0831 & 0.1974 \\      
        \hline
        & A29 & A30 & A31 & A32 & \multicolumn{2}{c|}{pooled}\\
        \hline
        min a-DCF & 0.0407 & 0.0783 & 0.0767 & 0.0683 & \multicolumn{2}{c|}{0.0756} \\
        \hline
    \end{tabular}
    }
\end{table}

The final results of the evaluation set system fusion submitted for the Track 2 open challenge are shown in Table 6. With the enhancement of self-supervised pre-training models and the strong performance of the speaker verification model, the detection performance of various spoofing algorithms has improved significantly, achieving a final min a-DCF of 0.0756.

\section{Conclusion}
In this paper, we present our system for the ASVspoof5 challenge. For Track 1, we utilized handcrafted features and self-supervised speech representations to address the closed and open conditions, respectively. Multiple systems were trained on an augmented dataset to effectively detect spoof attacks under various adversarial conditions. Our system achieved a minDCF of 0.3948 and an EER of 14.33\% in the closed condition, and a minDCF of 0.0750 and an EER of 2.59\% in the open condition, demonstrating robustness under adversarial conditions.
For Track 2, we implemented a two-stage ASV system, employing the QMF method to calibrate the system score. By integrating the CM system from Track 1 with the ASV system in a cascaded manner, we achieved a min-aDCF of 0.2814 in the closed condition and 0.0756 in the open condition, indicating superior performance in the SASV system.
%This template can be found on the conference website
%$<$https://www.asvspoof.org/workshop5$>$.

\clearpage
\bibliographystyle{IEEEbib}
\bibliography{ref}

\begin{thebibliography}{10}

\bibitem{1}
Edresson Casanova, Julian Weber, Christopher~D Shulby, et~al.,
\newblock ``Yourtts: Towards zero-sho multi-speaker tts and zero-shot voice conversion for everyone,''
\newblock in {\em International Conference on Machine Learning}, 2022, pp. 2709--2720.

\bibitem{2}
Zhizheng Wu, Tomi Kinnunen, Nicholas Evans, et~al.,
\newblock ``Asvspoof 2015: the first automatic speaker verification spoofing and countermeasures challenge,''
\newblock in {\em Interspeech 2015}, 2015, pp. 2037--2041.

\bibitem{3}
Tomi Kinnunen, Md. Sahidullah, Hector Delgado, et~al.,
\newblock ``The asvspoof 2017 challenge: Assessing the limits of replay spoofing attack detection,''
\newblock in {\em Interspeech 2017}, 2017, pp. 2--6.

\bibitem{4}
Massimiliano Todisco, Xin Wang, Ville Vestman, et~al.,
\newblock ``Asvspoof 2019: Future horizons in spoofed and fake audio detection,''
\newblock in {\em Interspeech 2019}, 2019, pp. 1008--1012.

\bibitem{5}
Junichi Yamagishi, Xin Wang, Massimiliano Todisco, et~al.,
\newblock ``Asvspoof 2021: accelerating progress in spoofed and deepfake speech detection,''
\newblock in {\em ASVspoof 2021 Workshop}, 2021, pp. 47--54.

\bibitem{6}
Jiangyan Yi, Jianhua Tao, Ruibo Fu, et~al.,
\newblock ``{ADD} 2023: The second audio deepfake detection challenge,''
\newblock in {\em IJCAI Workshop on Deepfake Audio Detection and Analysis}, 2023.

\bibitem{25}
Jee weon Jung, Hemlata Tak, Hye jin Shim, Hee-Soo Heo, Bong-Jin Lee, Soo-Whan Chung, Ha~jin Yu, Nicholas~WD Evans, and Tomi~H Kinnunen,
\newblock ``Sasv 2022: The first spoofing-aware speaker verification challenge,''
\newblock in {\em Proc. Interspeech}, 2022, vol. 2022, pp. 2893--2897.

\bibitem{7}
Haochen Wu, Zhuhai Li, Luzhen Xu, et~al.,
\newblock ``The {USTC-NERCSLIP} system for the {T}rack 1.2 of {A}udio {D}eepfake {D}etection ({ADD} 2023) challenge,''
\newblock in {\em IJCAI Workshop on Deepfake Audio Detection and Analysis}, 2023.

\bibitem{8}
Hemlata Tak, Massimiliano Todisco, Xin Wang, et~al.,
\newblock ``Automatic speaker verification spoofing and deepfake detection using wav2vec2.0 and data augmentation,''
\newblock in {\em Odyssey}, 2022, pp. 112--119.

\bibitem{9}
Haochen Wu, Jie Zhang, Zhentao Zhang, et~al.,
\newblock ``Robust spoof speech detection based on multi-scale feature aggregation and dynamic convolution,''
\newblock in {\em ICASSP}, 2024, pp. 10156--10160.

\bibitem{10}
Rongjie Huang, Chenye Cui, Feiyang Chen, et~al.,
\newblock ``Singgan: {G}enerative adversarial network for high-fidelity singing voice generation,''
\newblock in {\em ACM Multimedia 2022}, 2022, pp. 2525--2535.

\bibitem{11}
Hemlata Tak, Madhu Kamble, Jose Patino, et~al.,
\newblock ``Rawboost: A raw data boosting and augmentation method applied to automatic speaker verification anti-spoofing,''
\newblock in {\em ICASSP}, 2022, pp. 6382--6386.

\bibitem{12}
Ariel Cohen, Inbal Rimon, Eran Aflalo, and Haim Permuter,
\newblock ``A study on data augmentation in voice anti-spoofing,''
\newblock {\em Speech Communication}, vol. 141, pp. 56--67, 2022.

\bibitem{13}
Xu~Li, Na~Li, Chao Weng, et~al.,
\newblock ``Replay and synthetic speech detection with res2net architecture,''
\newblock in {\em ICASSP}, 2021, pp. 6354--6358.

\bibitem{14}
Haochen Wu, Jie Zhang, Zhentao Zhang, et~al.,
\newblock ``Robust spoof speech detection based on multi-scale feature aggregation and dynamic convolution,''
\newblock in {\em ICASSP}, 2024, pp. 10156--10160.

\bibitem{15}
You Zhang, Fei Jiang, and Zhiyao Duan,
\newblock ``One-class learning towards synthetic voice spoofing detection,''
\newblock {\em IEEE Signal Processing Letters}, vol. 28, pp. 937--941, 2021.

\bibitem{16}
Lixiang Li, Xiaopeng Xue, Haipeng Peng, et~al.,
\newblock ``Improved one-class learning for voice spoofing detection,''
\newblock in {\em 2023 Asia pacific signal and information processing association annual summit and conference}, 2023, pp. 1978--1983.

\bibitem{43}
Xin Wang, H{\'e}ctor Delgado, Hemlata Tak, Jee-weon Jung, Hye-jin Shim, Massimiliano Todisco, Ivan Kukanov, Xuechen Liu, Md~Sahidullah, Tomi Kinnunen, Nicholas Evans, Kong~Aik Lee, and Junichi Yamagishi,
\newblock ``{ASVspoof 5}: {Crowdsourced} speech data, deepfakes, and adversarial attacks at scale,''
\newblock in {\em ASVspoof Workshop 2024 (accepted)}, 2024.

\bibitem{18}
D.~Snyder, G.~Chen, and D.~Povey,
\newblock ``{MUSAN}: {A} music, speech, and noise corpus,''
\newblock in {\em arXiv preprint arXiv:1510.08484}, 2015.

\bibitem{19}
T.~Ko, V.~Peddinti, D.~Povey, et~al.,
\newblock ``A study on data augmentation of reverberant speech for robust speech recognition,''
\newblock in {\em ICASSP}, 2017, pp. 5220--5224.

\bibitem{26}
R~Huang, MWY Lam, J~Wang, D~Su, D~Yu, Y~Ren, and Z~Zhao,
\newblock ``Fastdiff: A fast conditional diffusion model for high-quality speech synthesis,''
\newblock in {\em IJCAI International Joint Conference on Artificial Intelligence}. IJCAI: International Joint Conferences on Artificial Intelligence Organization, 2022, pp. 4157--4163.

\bibitem{27}
Yuanjun Lv, Hai Li, Ying Yan, Junhui Liu, Danming Xie, and Lei Xie,
\newblock ``Freev: Free lunch for vocoders through pseudo inversed mel filter,''
\newblock in {\em Proc. Interspeech}, vol. 2024.

\bibitem{28}
Jungil Kong, Jaehyeon Kim, and Jaekyoung Bae,
\newblock ``Hifi-gan: Generative adversarial networks for efficient and high fidelity speech synthesis,''
\newblock {\em Advances in neural information processing systems}, vol. 33, pp. 17022--17033, 2020.

\bibitem{29}
Jaehyeon Kim, Jungil Kong, and Juhee Son,
\newblock ``Conditional variational autoencoder with adversarial learning for end-to-end text-to-speech,''
\newblock in {\em International Conference on Machine Learning}. PMLR, 2021, pp. 5530--5540.

\bibitem{30}
J~Chung, A~Nagrani, and A~Zisserman,
\newblock ``Voxceleb2: Deep speaker recognition,''
\newblock {\em Interspeech 2018}, 2018.

\bibitem{35}
Hemlata Tak, Jose Patino, Massimiliano Todisco, Andreas Nautsch, Nicholas Evans, and Anthony Larcher,
\newblock ``End-to-end anti-spoofing with rawnet2,''
\newblock in {\em ICASSP 2021-2021 IEEE International Conference on Acoustics, Speech and Signal Processing (ICASSP)}. IEEE, 2021, pp. 6369--6373.

\bibitem{33}
Penghui Wen, Kun Hu, Wenxi Yue, Sen Zhang, Wanlei Zhou, and Zhiyong Wang,
\newblock ``Robust audio anti-spoofing with fusion-reconstruction learning on multi-order spectrograms,''
\newblock in {\em Interspeech 2023}, pp. 271--275.

\bibitem{31}
Feng Wang, Jian Cheng, Weiyang Liu, and Haijun Liu,
\newblock ``Additive margin softmax for face verification,''
\newblock {\em IEEE Signal Processing Letters}, vol. 25, no. 7, pp. 926--930, 2018.

\bibitem{32}
Yifan Sun, Changmao Cheng, Yuhan Zhang, Chi Zhang, Liang Zheng, Zhongdao Wang, and Yichen Wei,
\newblock ``Circle loss: A unified perspective of pair similarity optimization,''
\newblock in {\em Proceedings of the IEEE/CVF conference on computer vision and pattern recognition}, 2020, pp. 6398--6407.

\bibitem{34}
Michael~JD Powell,
\newblock ``A view of algorithms for optimization without derivatives,''
\newblock {\em Mathematics Today-Bulletin of the Institute of Mathematics and its Applications}, vol. 43, no. 5, pp. 170--174, 2007.

\bibitem{20}
Alexei Baevski, Henry Zhou, Abdelrahman Mohamed, et~al.,
\newblock ``Wav2vec2.0: {A} framework for self-supervised learning of speech representations,''
\newblock in {\em Adv. Neural Inf. Process. Syst.(NeurIPS)}, 2020, pp. 12449--12460.

\bibitem{21}
Xin Wang and Junichi Yamagishi,
\newblock ``{A Comparative Study on Recent Neural Spoofing Countermeasures for Synthetic Speech Detection},''
\newblock in {\em Interspeech 2021}, 2021, pp. 4259--4263.

\bibitem{22}
Jee weon Jung, Hee-Soo Heo, Hemlata Tak, et~al.,
\newblock ``{AASIST}: {A}udio anti-spoofing using integrated spectro-temporal graph attention networks,''
\newblock in {\em ICASSP}, 2022, pp. 6367--6371.

\bibitem{23}
Shengyu Peng, Wu~Guo, Haochen Wu, et~al.,
\newblock ``{Fine-tune Pre-Trained Models with Multi-Level Feature Fusion for Speaker Verification},''
\newblock in {\em Interspeech 2024}, 2024.

\bibitem{36}
Kaiming He, Xiangyu Zhang, Shaoqing Ren, and Jian Sun,
\newblock ``Deep residual learning for image recognition,''
\newblock in {\em Proceedings of the IEEE conference on computer vision and pattern recognition}, 2016, pp. 770--778.

\bibitem{37}
Miao Zhao, Yufeng Ma, Yiwei Ding, Yu~Zheng, Min Liu, and Minqiang Xu,
\newblock ``Multi-query multi-head attention pooling and inter-topk penalty for speaker verification,''
\newblock in {\em ICASSP 2022-2022 IEEE International Conference on Acoustics, Speech and Signal Processing (ICASSP)}. IEEE, 2022, pp. 6737--6741.

\bibitem{39}
Jiankang Deng, Jia Guo, Niannan Xue, and Stefanos Zafeiriou,
\newblock ``Arcface: Additive angular margin loss for deep face recognition,''
\newblock in {\em Proceedings of the IEEE/CVF conference on computer vision and pattern recognition}, 2019, pp. 4690--4699.

\bibitem{38}
Jiankang Deng, Jia Guo, Tongliang Liu, Mingming Gong, and Stefanos Zafeiriou,
\newblock ``Sub-center arcface: Boosting face recognition by large-scale noisy web faces,''
\newblock in {\em Computer Vision--ECCV 2020: 16th European Conference, Glasgow, UK, August 23--28, 2020, Proceedings, Part XI 16}. Springer, 2020, pp. 741--757.

\bibitem{40}
Jenthe Thienpondt, Brecht Desplanques, and Kris Demuynck,
\newblock ``The idlab voxceleb speaker recognition challenge 2020 system description,''
\newblock {\em arXiv preprint arXiv:2010.12468}, 2020.

\bibitem{41}
Yu~Zheng, Jinghan Peng, Yihao Chen, Yajun Zhang, Jialong Wang, Min Liu, and Minqiang Xu,
\newblock ``The speakin speaker verification system for far-field speaker verification challenge 2022,''
\newblock {\em arXiv preprint arXiv:2209.11625}, 2022.

\end{thebibliography}
\end{document}